\documentclass[a4paper, 10pt]{amsproc}
\usepackage{amssymb}
\usepackage{amscd}
\usepackage[dvips]{graphicx}
\usepackage[all,cmtip]{xy}
\usepackage[inline]{enumitem}
\usepackage{xcolor}
\usepackage{cite}
\usepackage[hidelinks]{hyperref}

\makeatletter
\renewcommand*{\eqref}[1]{%
  \hyperref[{#1}]{\textup{\tagform@{\ref*{#1}}}}%
}
\makeatother

\newtheorem{thmA}{Main result}
\newtheorem{theorem}{Theorem}
\newtheorem{proposition}[theorem]{Proposition}
\newtheorem{lemma}[theorem]{Lemma}
\newtheorem{corollary}[theorem]{Corollary}
\theoremstyle{definition}
\newtheorem{example}{Example}
\newtheorem{definition}{Definition}
\newtheorem{remark}{Remark}

\newcommand{\R}{\mathbb{R}}

\newcommand{\ddim}{\mathrm{ddim\;}}
\newcommand{\dind}{\mathrm{dind\;}}
\newcommand{\corank}{\mathrm{corank\;}}
\newcommand{\g}{\mathfrak{g}}
\newcommand{\kk}{\mathfrak{k}}

\newcommand{\uu}{\mathfrak{u}}

\newcommand{\so}{\mathfrak{so}}
\newcommand{\spp}{\mathfrak{sp}}

\newcommand{\ad}{\mathrm{ad}}
\DeclareMathOperator{\Ad}{\mathrm{Ad}}
\DeclareMathOperator{\Span}{\mathrm{span\,}}
\DeclareMathOperator{\diag}{\mathrm{diag}}

\DeclareMathOperator{\rank}{\mathrm{rank}}
 \DeclareMathOperator{\pr}{\mathrm{pr}}
\DeclareMathOperator{\tr}{\mathrm{tr}}

\allowdisplaybreaks

\begin{document}

\title[Integrable systems associated to the filtrations of Lie algebras]{Integrable systems associated to the filtrations of Lie algebras}

\author{Bo\v zidar Jovanovi\'c, Tijana \v Sukilovi\' c, Srdjan Vukmirovi\'c}

\address{B.J.: Mathematical Institute, Serbian Academy of Sciences and
Arts, Kneza Mihaila 36, 11000 Belgrade, Serbia}
\email{bozaj@mi.sanu.ac.rs}

\address{ T.\v{S}, S.V:  Faculty of Mathematics, University of Belgrade, Studentski trg 16, 11000 Belgrade, Serbia}
\email{vsrdjan@matf.bg.ac.rs, tijana@matf.bg.ac.rs}



\begin{abstract}
In 1983 Bogoyavlenski conjectured that if the Euler equations on a Lie algebra $\g_0$ are integrable, then their certain extensions to semisimple lie algebras $\g$ related to the filtrations of Lie algebras
$\g_0\subset\g_1\subset\g_2\dots\subset\g_{n-1}\subset \g_n=\g$ are integrable as well.
In particular, by taking $\g_0=\{0\}$ and natural filtrations of $\so(n)$ and $\uu(n)$, we have
Gel'fand-Cetlin integrable systems. We proved the conjecture for filtrations of compact Lie algebras $\g$: the system are integrable in a noncommutative sense by means of polynomial integrals.
Various constructions of complete commutative polynomial integrals for the system are also given.
\end{abstract}

\maketitle

\section{Introduction}

Let $G$ be a compact connected Lie group.
Consider a chain of
connected compact Lie subgroups
\begin{align*}
G_0 \subset G_1 \subset G_2 \subset \dots \subset G_{n-1}\subset
G_n=G
\end{align*}
and the corresponding filtration of the Lie algebra $\g=Lie(G)$
\begin{align}\label{filtration}
\g_0\subset\g_1\subset\g_2\dots\subset\g_{n-1}\subset \g_n=\g.
\end{align}

We study integrable Euler equations related to the filtration
\eqref{filtration}. One can consider non compact Lie algebras as
well. In fact, one of the first contribution is given by Trofimov,
who constructed integrable systems on Borel subgroups of complex
semisimple Lie algebras (see~\cite{Tr}). Later,
Bogoyavlenski~\cite{B1} considered filtration of semisimple Lie
algebras, such that the restrictions of the Killing form to
$\g_i$, $i=0,\dots,n$ are non-degenerate. We restrict
ourself to the compact case and a generalization of
Gel'fand-Cetlin systems on Lie algebras $\so(n)$ and $\uu(n)$ given by filtrations
\eqref{so(n)} and~\eqref{u(n)} below in order to insure compact invariant manifolds of the flows.
Similar statements can be formulated for reductive groups as well.

Fix an invariant scalar product
$\langle\,\cdot\,,\,\cdot\,\rangle$ on $\g$ and denote the
restrictions of $\langle\,\cdot\,,\,\cdot\,\rangle$ to $\g_i$
also by $\langle\,\cdot\,,\,\cdot\,\rangle$.
By the use of $\langle\,\cdot\,,\,\cdot\,\rangle$, we identify
$\g\cong\g^*$ and
$\g_i\cong\g^*_i$, $i=0,\dots,n$.
Let $\mathfrak p_i$ be the orthogonal complement of $\g_{i-1}$ in $\g_i$, $\mathfrak p_0=\g_0$ and
$\pr_{\mathfrak p_i}$ and $\pr_{\g_i}$ the orthogonal
projections onto $\mathfrak p_i$ and $\g_i$, respectively.
For $x\in\g$, we denote
\begin{align*}
y_i=\pr_{\mathfrak p_i}(x), \qquad x_i=y_0+y_1+\dots+y_i=\pr_{\g_i}(x), \qquad i=0,\dots,n.
\end{align*}

The Euler equations
\begin{align}
\dot x=[x, \omega], \qquad \omega=A(x) \label{euler}
\end{align}
 associated with a symmetric positive operator of the form
\begin{align}\label{operatorBo}
A(x)=A_0(y_0)+\sum_{i=1}^n s_i y_i, \quad A_0:\g_0\to\g_0, \quad s_i\in\R, \quad  i=1,\dots,n
\end{align}
were studied by Bogoyavlenski~\cite{B1}.
The equations are Hamiltonian with respect to the Lie--Poisson bracket\footnote{The gradient is determined by an invariant metric: $df(\xi)=\langle \nabla f(x),\xi\rangle$. Also, to simplify notation, the Lie brackets, the Lie-Poisson brackets and the gradients of the functions on $\g_i$ will be denoted by the same symbols as on $\g$, $i=1,\dots,n$.}
\begin{align}\label{LP}
\{f,g\}\vert_x=-\langle x,[\nabla f(x),\nabla g(x)]\rangle
\end{align}
and the Hamiltonian function $H(x)=\frac12\langle x,\omega\rangle=\frac12\langle A(x),x\rangle$.

Due to the relations
\begin{align*}
[\mathfrak p_i,\mathfrak p_j]\subset \mathfrak p_j, \qquad  0\le
i<j\le n,
\end{align*}
the Euler equations~\eqref{euler} can be rewritten into the form
\begin{align}
\dot y_0 =& [y_0,A_0(y_0)], \label{euler-0}\\
\dot y_i =& [y_i,A_0(y_0)-s_i y_0+(s_1-s_i) y_1+\dots+(s_{i-1}-s_i)
y_{i-1}], \quad i=1,\dots,n. \label{euler-i}
\end{align}

Specially, if $\g_0=\{0\}$ is a trivial Lie algebra, we
have $y_1=const$ and the components of $y_2$ are elementary
functions of the time $t$.

The system~\eqref{euler-0},~\eqref{euler-i} has obvious family of
polynomial first integrals
\begin{align}\label{thimm}
\mathcal I=\mathcal I_1+\mathcal I_2+\dots+\mathcal I_n,
\end{align}
where $\mathcal I_i$ are invariants $\R[\g_i]^{G_i}$
lifted to $\g$ along the projection to ${\g_i}$: $ \mathcal
I_i=\pr_{\g_i}^*\R[\g_i]^{G_i}$, $i=1,\dots,n$. According to the
following (quite simple, but important) lemma, it is clear that
$\mathcal I$ is a commutative algebra with respect to the
Lie-Poisson bracket~\eqref{LP}.

\begin{lemma} [\cite{B1, Tr, Th}] \label{trof}
If $f$ and $g$ Lie--Poisson commute on $\g_i$, then their lifts
$\tilde f=\pr^*_{\g_i} f$ and $\tilde g=\pr^*_{\g_i} g$ Lie--Poisson commute on
$\g$.
\end{lemma}

Bogoyavlenski found also a large class of additional first integrals obtained
by the translations of invariants of $\g_i$ along the subalgebra $\g_{i-1}$
\begin{align}\label{intB}
p_{\alpha,\beta}(x)=p(\alpha x_i+\beta y_i), \quad p\in \R[\g_i]^{G_i},  \quad i=1,\dots,n, \quad \alpha,\beta\in\R
\end{align}
and conjectured that the equations~\eqref{euler-0},
\eqref{euler-i} are completely integrable if this is true for the
Euler equations~\eqref{euler-0} (see~\cite{B1}). In~\cite{Mik}, Mikityuk proved that Bogoyavlenski's
integrals~\eqref{intB} imply complete commutative integrability in the case when $(\g_i,\g_{i-1})$ are symmetric pairs,
that is when
\begin{align*}
[\mathfrak p_i,\mathfrak p_i]\subset \g_{i-1}, \qquad i=1,\dots,n.
\end{align*}

On the other hand, Thimm
used chains of subalgebras~\eqref{filtration} in studying the
integrability of geodesic flows on homogeneous spaces (see~\cite{Th}).
He proved that integrals~\eqref{thimm} form a complete
commutative algebras on the Lie algebras $\so(n)$ and $\uu(n)$, with
respect to the natural filtrations
\begin{align}\label{so(n)}
\so(2)\subset \so(3)\subset \dots \subset \so(n-1)\subset \so(n)
\end{align}
and
\begin{align}\label{u(n)}
 \uu(1)\subset \uu(2) \subset \dots \subset \uu(n-1)\subset \uu(n),
\end{align}
respectively.

After~\cite{GS}, the corresponding integrable systems are refereed
as Gel'fand-Cetlin systems on $\so(n)$ and $\uu(n)$. Namely, Gel'fand
and Cetlin constructed canonical bases for a finite-dimensional
representation of the orthogonal and unitary groups by the
decomposition of the representation by a chain of subgroups~\cite{GC1, GC2}.
The corresponding integrable systems on the adjoint
orbits with integrals $\mathcal I$ can be seen as a symplectic
geometric version of the
Gelfand-Cetlin construction~\cite{GS}. Also, Thimm's examples
motivated Guillemin and Sternberg to introduce an important notion
of multiplicity free Hamiltonain actions~\cite{GS1, GS2}.
The construction is also used in the study of integrability of geodesic flows on homogeneous spaces
and bi-quotients of Lie groups (see~\cite{Th, BJ1, BJ3, Baz}).
The nonholonimic systems on compact Lie groups $G$ with left invariant metrics defined by
the Hamiltonians of the form $H=\frac12\langle A(x),x\rangle$, where $A$ is given by \eqref{operatorBo},
and left-invariant constraints are studied in \cite{JoNon}.

In this paper we prove complete integrability in a
noncommutative sense of the system~\eqref{euler-0},~\eqref{euler-i} and calculate the dimension of invariant isotropic tori $\delta$
(Theorem~\ref{th1}, Section~\ref{sec3}):
\begin{thmA}\label{glavna1}
Assume that the Euler equations~\eqref{euler-0}
are integrable by polynomial integrals. Then the Euler equations~\eqref{euler-0},
\eqref{euler-i} are completely integrable in a noncommutative sense by means of polynomial integrals as well.
\end{thmA}

Concerning dynamics, noncommutative (or superintegrability) is a stronger characteristic then the usual commutative (or Liouville) integrability. The system is solvable by quadratures, regular compact invariant manifolds are isotropic tori, and there exist an appropriate action--angle coordinates in which the dynamics is linearized~\cite{N, MF}.
It implies the Liouville integrability, at least by means of smooth functions: invariant isotropic tori can be always organized into resonant Lagrangian tori~\cite{BJ2}.

According to the Mischenko-Fomenko conjecture, a natural algebraic problem is a construction of a complete commutative set of polynomial integrals. The problem can be formulated in terms of pairs $(G_{i-1},G_i)$:  we need to construct
\begin{align}\label{bUSLOV}
\mathbf b(\g_i,\g_{i-1})=\frac12(\dim\mathfrak p_i+\rank\g_i-\rank\g_{i-1})
\end{align}
mutually independent commuting $\Ad_{G_{i-1}}$--invariants on $\g_i$  that are independent from the polynomials on $\g_{i-1}$
(Corollary~\ref{cor1}, Theorem~\ref{th2}, Section~\ref{sec4}).
 The basic examples are pairs $(G_{i-1},G_i)$ where $G_{i-1}$ is a multiplicity free or an almost multiplicity free subgroup of $G_i$
(see Examples \ref{ex1} and \ref{ex2}, Section~\ref{sec4}).

In Section~\ref{sec6} we adopt various constructions of commutative polynomials on $\g_i$ to provide complete commutative polynomial integrals for the system~\eqref{euler-0},
\eqref{euler-i}. 
When $G_{i-1}$ is an isotropy subgroup of $a_i\in \mathfrak g_i$, one can use the Mishchenko-Fomenko shifting of argument method in both cases: when $a_i$ is regular (see~\cite{MF1, Bo2}),
but also a singular element of $\g_i$ (see Theorem~\ref{singMF}).
Further, as an example of a variation of Mikityuk's construction  and a complement to~\eqref{so(n)} and~\eqref{u(n)},
a complete commutative sets of polynomials for the filtrations
\begin{align*}
\spp(k_0)\subset \spp(k_1)\subset \dots\subset \spp(k_n), \qquad k_0 < k_1 <\dots <k_n
\end{align*}
are given (Proposition~\ref{sp-thm}, see also~\cite{Ha}).

Also, in subsection~\ref{poredjenje}
we estimate the number of independent Bogoyavlenski's integrals~\eqref{intB} (Theorem~\ref{GCM}).

\begin{thmA}
Under certain generic assumptions, among  Bogoyavlenski's integrals~\eqref{intB} there are
at least $\mathbf b(\g_i,\g_{i-1})$ mutually independent polynomials (see \eqref{bUSLOV}) that are independent from the polynomials on $\g_{i-1}$.
\end{thmA}

Although the polynomials~\eqref{intB} do not commute in general,
we provide examples of complete commutative sets of integrals obtained by Bogoyavlenski's method (Example~\ref{exCom}).
Finally, in Theorems~\ref{pod1},~\ref{pod2},~\ref{pod3} we recall the results obtained in~\cite{Jo, Bo2, Re, DGJ}, which solve the problem in several interesting cases.

For the completeness of the exposition, we begin the presentation by briefly recalling on the concept of noncommutative integrability.

\section{Complete algebras of functions on Poisson manifolds}\label{sec2}

Let $(M,\Lambda)$ be a Poisson manifold. The Poisson bracket of two smooth functions is defined by the use of the Poisson tensor $\Lambda$ as usual $\{f,g\}\vert_x =\Lambda_x(df(x),dg(x))$, giving the
Lie algebra structure to $C^\infty(M)$.
Let $x=(x_1,\dots,x_m)$ be local coordinates on $M$. Let $f$ and $g$ be the first integrals of the Hamiltonian equations with the Hamiltonian $H$:
\begin{align}\label{HamEq}
\dot x=X_H=\sum_i\{x_i,H\}\frac{\partial}{\partial x_i}=\sum_{i,j} \Lambda_{ij} \frac{\partial H}{\partial x_j}\frac{\partial}{\partial x_i},
\end{align}
that is $\{f,H\}=\{g,H\}=0$.
Then, due to the Jacobi identity, their Poisson bracket $\{f,g\}$ is also the first integral: $\{\{f,g\},H\}=0$.

Therefore we can consider the Lie algebra $\mathcal F$ of first integrals.
Let $F_x=\{df\,\vert\, f\in\mathcal F\}$ be a subspace of $T^*_xM$ spanned by differentials of the functions from $\mathcal F$ at $x\in M$.
Assume that the dimensions of $F_x$ and $\dim\ker\Lambda_x\vert_{F_x}$ are constant on an open dense set $U$ of $M$.
The corresponding dimensions are denoted by $\ddim\mathcal F$ (\emph{differential dimension} of $\mathcal F$)
and $\dind\mathcal F$ (\emph{differential index} of $\mathcal F$), respectively.

\begin{remark}\label{primedba1}
The numbers $l=\ddim\mathcal F$ and $k=\dind\mathcal F$ have a clear geometrical meaning.
Let $y_1,y_2,\dots,y_l\in\mathcal F$ be independent functions within a domain $V\subset U$, such that
\begin{align*}
\mathbf F\colon V\to W\subset \R^l, \quad \mathbf F(x)=(y_1(x),\dots,y_l(x))
\end{align*}
is a submersion and
\begin{align}\label{newPB}
\{y_i,y_j\}=a_{ij}(y_1,\dots,y_l),
\end{align}
where $a_{ij}$ are smooth functions on $W$. Then \eqref{newPB} defines a Poisson structure on $W$ of a constant corank $k$.
\end{remark}

$\mathcal F$ is a \emph{complete algebra} (or a \emph{complete algebra at} $x$, $x\in U$) if
\begin{align*}
\ddim\mathcal F+\dind\mathcal F=\dim M+\corank \Lambda,
\end{align*}
or, equivalently,  if
\begin{align}\label{uslovKompletnost}
F_x^\Lambda=\{\xi\in T^*_x M\,\vert\, \Lambda_x(F_x,\xi)=0\}\subset F_x, \qquad x\in U.
\end{align}

{Similarly, if $\mathcal F$ is an arbitrary Lie subalgebra of $C^\infty(M)$ and $\mathcal F_0\subset \mathcal F$ its subalgebra, we say that $\mathcal F_0$ is a
\emph{complete subalgebra of} $\mathcal F$ if
\begin{align*}
\ddim\mathcal F_0+\dind\mathcal F_0=\ddim\mathcal F+\dind\mathcal F.
\end{align*}
In particular, $\mathcal F$ is a complete algebra on $M$ if and only if it is a complete subalgebra of $C^\infty(M)$.}
If $\mathcal F$ is a complete algebra on $M$ then $\mathcal F\vert_\mathcal S$ (the restrictions of the functions to $\mathcal S$) will be a complete algebra on a generic symplectic leaf $\mathcal S\subset M$.
Specially, we may be interested in the competentness of $\mathcal F$ not on $M$ but on  a particular, regular or singular, symplectic leaf $\mathcal S_0$ (see Lemma~\ref{potpunost} below).
Then the condition~\eqref{uslovKompletnost} is slightly changed:
$\mathcal F\vert_{\mathcal S_0}$ is complete
\begin{align*}
\ddim(\mathcal F\vert_{\mathcal S_0})+\dind(\mathcal F\vert_{\mathcal S_0})=\dim\mathcal S_0,
\end{align*}
if
$F_x^\Lambda=\{\xi\in T^*_x M\,\vert\, \Lambda_x(F_x,\xi)=0\}\subset F_x+\ker\Lambda_x$,
for a generic $x\in \mathcal S_0$.

The Hamiltonian system~\eqref{HamEq} is called \emph{completely integrable in a noncommutative sense} (or \emph{superintegrable}), if it has
a complete algebra $\mathcal F$ of first integrals.
The integrable system is solvable by quadratures (at least locally),
the regular compact connected components of the level sets determined by functions in $\mathcal F$ are $\delta$--dimensional (isotropic considered on
the symplecic leaves of $\Lambda$) tori ($\delta=\dim M-\ddim\mathcal F=\dind\mathcal F-\corank\Lambda$), and there exist an appropriate action--angle coordinates in which the dynamics is linearized~\cite{N, MF}.

When $\mathcal F$ is a complete commutative algebra,
\begin{align*}
\ddim\mathcal F=\dind\mathcal F=\mathbf a(M)=\frac12\big(\dim M+\corank \Lambda\big),
\end{align*}
we have the usual Liouville integrability -- regular compact invariant level sets of the functions from $\mathcal F$ are
$\delta_0=\dim M-\mathbf a(M)$ dimensional (Lagrangian on the symplectic leaves) tori.

\begin{remark}\label{primedba2}
Note that $\mathbf a(M)$ is the maximal number of Poisson commuting independent functions on $M$.
For an arbitrary $\mathcal F$ we have the inequality
\begin{align}\label{nejednakost}
\ddim\mathcal F+\dind\mathcal F \le 2\mathbf a(M),
\end{align}
(or, concerning Remark~\ref{primedba1}, $\mathbf a(W)\le \mathbf a(M)$) and equality holds if and only if $\mathcal F$ is a complete algebra.
\end{remark}

Mishchenko and Fomenko stated the conjecture that noncommutative integrability implies the Liouville integrability by means of an algebra of integrals that belong to the same functional class as the original one~\cite{MF}.
Note that in the case of noncommutative integrability trajectories of~\eqref{HamEq} belong to the tori of dimension $\delta<\delta_0=\dim M-\mathbf a(M)$, that is, $\delta_0$--dimensional Lagrangian invariant tori are resonant: they are filled with
$(\delta_0-\delta)$--parametric family of $\delta$-dimensional invariant tori. In a smooth category the problem is easy to solve: we can always semi--locally reorganize isotropic toric foliation into the Lagrangian toric foliation  (see~\cite{BJ2}\footnote{In~\cite{BJ2} the symplectic case is considered, but the proof can be easily modified to the Poisson case.}).
The polynomial Mishchenko-Fomenko conjecture says that one can find independent commuting functions $p_1,\dots,p_\mathbf a$, that are polynomials in functions from $\mathcal F$.
The conjecture is solved for finite dimensional Lie algebras $\mathcal F$ (see~\cite{Bo, Sa}).  From a point of view of the dynamics, noncommutative integrability is stronger then the Liouville one since Lagrangian tori are resonant and not an intrinsic property of the system.

\section{Polynomial noncommutative integrability}\label{sec3}

Suppose that the Euler equations~\eqref{euler-0} are completely
integrable by means of a complete algebra $\mathfrak A_0$ of
polynomial integrals,
\begin{align*}
\ddim\mathfrak A_0+\dind\mathfrak A_0=\dim\g_0+\rank\g_0,
\end{align*}
and let $\delta_0=\dind\mathfrak A_0-\rank\g_0$ be the dimension
of generic invariant tori.

Let $\mathfrak A_i$ be the algebra $\R[\g_i]^{G_{i-1}}$
of $\Ad_{G_{i-1}}$--invariant polynomials on $\g_i$. Consider the lift of
algebras $\mathfrak A_i$ to the Lie algebra $\g$:
\begin{align*}
\mathcal A_i=\pr_{\g_i}^*\mathfrak A_i, \qquad i=0,\dots,n.
\end{align*}
In particular, since the invariants on $\g_i$ belong to $\mathfrak
A_i$, we have $\mathcal I_i\subset\mathcal A_i$.

\begin{theorem}\label{th1}
The system~\eqref{euler-0},~\eqref{euler-i} is completely
integrable with a complete set of polynomial integrals
\begin{align*}
\mathcal A=\mathcal A_0+ \mathcal A_1+\dots+\mathcal A_n.
\end{align*}

A generic motion is a quasi-periodic winding over
\begin{align*}
\delta=\delta_0+\rank\g_0-\rank\g+\sum_{i=1}^n
\dim\pr_{\mathfrak p_i}(\g_i(x_i))
\end{align*}
dimensional invariant tori determined by the integrals $\mathcal A$. Here we take
generic elements $x_i\in\g_i$, $i=1,\dots,n$.\footnote{$\g_{l}(x_{k})$ denotes the
isotropy algebra of $x_k\in\g_k$ within $\g_l$:
\begin{align*}
\g_l(x_k)=\{\xi\in\g_l\,\vert\,[\xi,x_k]=0\}, \qquad l\le k.
\end{align*}
Generic
means that the dimensions of
the isotropy algebras $\g_{i}(x_i)$ and $\g_{i-1}(x_i)$ are minimal.}
\end{theorem}

Recall that for a generic $x_i\in\g_i$, $\g_i(x_i)$ is a Cartan subalgebra in $\g$ that is spanned by the gradients of $\rank\g_i$ basic invariant polynomials in $\R[\g_i]^{G_i}$,
which also coincides with the kernel of the Lie-Poisson structure of $\g_i$ at $x_i$.

\begin{proof}
First, note that polynomials $\mathcal A_i$
($i>0$) are indeed integrals of the equations. The polynomial $p$
belongs to $\mathfrak A_i$, $i=1,\dots,n$, if and only if
\begin{align*}
\langle \nabla p(x_i),[\xi,x_i]\rangle=0, \quad \text{for all}
\quad \xi\in\g_{i-1}, \, x_i\in\g_i,
\end{align*}
which is equivalent to the Poisson commuting of $p$ with the lifting of all polynomials from $\R[\mathfrak g_{i-1}]$ to $\R[\g_i]$.

Let $\tilde p=\pr_{\g_{i}}^* p$. Then
\begin{align*}
\frac{d}{dt}\tilde p(x)=\langle \nabla p(x_i), \dot x_{i}\rangle
=\langle \nabla p(x_i), [x_i,A(x_{i})]\rangle= \langle \nabla
p(x_i), [x_i,A(x_{i-1})-s_i x_{i-1}]\rangle=0.
\end{align*}

Let $\mathcal O_i\subset\g_i$ be a generic $G_i$-adjoint orbit.
With a sign convention \eqref{LP},
the momentum mapping of $G_i$ adjoint action is simply the inclusion
mapping $\imath: \mathcal O_i\hookrightarrow \g_i$  multiplied by $-1$, while the
momentum mapping of the adjoint action of $G_{i-1}$ is
\begin{align*}
\Phi_{i-1}: \mathcal O_i \to \g_{i-1}, \qquad
\Phi_{i-1}=-\pr_{\g_{i-1}}\circ\,\imath.
\end{align*}

According to the Lemma~\ref{potpunost} on integrability related to Hamiltonian actions (see below),
the algebra $C^\infty_{G_{i-1}}(\mathcal
O_i)+\Phi^*_{i-1}(\R[\g_{i-1}])$ is a complete algebra on
$\mathcal O_i$, where $C^\infty_{G_{i-1}}(\mathcal O_i)$ is the
algebra of smooth $G_{i-1}$-invariant functions. Since generic
orbit of $\Ad_{G_{i-1}}$-action on $\g_i$ are separated by invariant
polynomials, we have that
$
\mathfrak A_i\vert_{\mathcal O_i}+\Phi^*_{i-1}(\R[\g_{i-1}])
$
is a complete polynomial algebra on $\mathcal O_i$.
Therefore
\begin{align}\label{complete_i}
\mathfrak A_i+\pr^*_{i-1}(\R[\g_{i-1}])
\end{align}
is a complete algebra on $\g_i$ (recall that the invariants $\R[\g_i]^{G_i}$ are contained in $\mathfrak A_i=\R[\g_i]^{G_{i-1}}$). Next, by induction using the item~\ref{potpunost_ii} of Lemma~\ref{potpunost}, we get that
$\mathcal A$ is a complete algebra of integrals on $\g$.

In order to determine the dimension of invariant tori,  note that
the dimension of invariant tori determined by the functions
$
\mathfrak A_i\vert_{\mathcal O_i}+\Phi^*_{i-1}(\R[\g_{i-1}])
$
on $\mathcal O_i$ equals (Lemma~\ref{potpunost}):
\begin{align*}
\delta_i =& \dind(\mathfrak A_i\vert_{\mathcal O_i}+\Phi^*_{i-1}(\R[\g_{i-1}]))=
\dind(\mathfrak A_i\vert_{\mathcal O_i})=\dind(\Phi^*_{i-1}(\R[\g_{i-1}])))\\
=&\dim \g_{i-1}(x_{i-1})-\dim \g_{i-1}(x_i)\\
=&\rank
\g_{i-1}-\dim(\g_{i-1}\cap \g_{i}(x_i))\\
=& \rank \g_{i-1}-\rank \g_i+\dim\pr_{\mathfrak
p_i}(\g_i(x_i)).
\end{align*}
Here $x_i\in\mathcal O_i$ is a generic element and
$x_{i-1}=\pr_{\g_{i-1}}(x_i)$.

Again, by induction using the item~\ref{potpunost_ii} of Lemma~\ref{potpunost}, we get the dimension of invariant tori:
\begin{align*}
\delta =& \delta_0+\delta_1+\dots+\delta_n \\
=& \delta_0+\rank \g_{0}-\rank \g_1+\dim\pr_{\mathfrak p_1}(\g_1(x_1))+\rank \g_{1}-\rank \g_2+\dim\pr_{\mathfrak p_2}(\g_2(x_2))\\
& +\dots+\rank \g_{n-1}-\rank \g_n+\dim\pr_{\mathfrak p_n}(\g_n(x_n))\\
=& \delta_0+\rank\g_0-\rank\g_n+\dim\pr_{\mathfrak p_1}(\g_1(x_1))+\dots+\dim\pr_{\mathfrak p_n}(\g_n(x_n)),
\end{align*}
for a generic $x_i\in\g_i$, $i=1,\dots,n$.
\end{proof}

Here we recall on the construction of collective integrable systems.
Consider the Hamiltonian action of a compact connected Lie group $K$ on the
symplectic manifold $M$ with the equivariant momentum mapping $\Phi: M\to \kk^*\cong \kk$.

Let $C^\infty_K(M)$ be the algebra of $K$--invariant functions.
According to the Noether theorem, $\{f,\tilde p\}_M=0$ for all $f\in C^\infty_K(M)$, $\tilde p=\Phi\circ p$, $p\in\R[\kk]$.
Also, if $p\in\R[\kk]^K$ is an invariant polynomial, then $\tilde p$ is $K$--invariant. In~\cite{BJ2} we proved the following quite simple but important statement.

\begin{lemma}[\cite{BJ2}]\label{potpunost}
\begin{enumerate}[wide, topsep=0pt, itemsep=1pt, labelwidth=!, labelindent=0pt, label=(\roman*)]
\item\label{potpunost_i}
The algebra of functions $C^\infty_K(M)+\Phi^*(\R[\kk])$ is complete on $M$:
\begin{align*}
\ddim(C^\infty_K(M)+\Phi^*(\R[\kk]))+\dind(C^\infty_K(M)+\Phi^*(\R[\kk]))=\dim M
\end{align*}
and the dimension of invariant regular isotropic tori is
\begin{align*}
\delta_0=\dind(C^\infty_K(M)+\Phi^*(\R[\kk]))=\dind C^\infty_K(M)=\dind\Phi^*(\R[\kk])=\dim K_\mu-\dim K_x,
\end{align*}
where $K_x$ and $K_\mu$ are isotropic subgroups of a generic $x\in M$ and $\mu=\Phi(x)\in\kk$.

\item\label{potpunost_ii} Let $\mathfrak F\subset \R[\kk]$ be complete on a generic adjoint orbit in the image of $M$,
\begin{align*}
\ddim(\mathfrak F\vert_\mathcal O)+\dind(\mathfrak F\vert_\mathcal O)=\dim\mathcal O,\qquad \mathcal O\subset \Phi(M),
\end{align*}
and let $\mathcal F=\Phi^*\mathfrak F$. Then $C^\infty_K(M)+\mathcal F$ is complete on $M$:
\begin{align*}
\ddim(C^\infty_K(M)+\mathcal F)+\dind(C^\infty_K(M)+\mathcal F)=\dim M
\end{align*}
and the dimension of invariant isotropic tori is
\begin{align*}
\delta_1=\dind(C^\infty_K(M)+\mathcal F)=\delta_0+\dind(\mathfrak F\vert_\mathcal O).
\end{align*}
\end{enumerate}
\end{lemma}

Note that all adjoint orbits in $\Phi(M)$ could be singular and then the completeness of $\mathfrak F$ on $\kk$ does not imply directly the completeness
of the restriction $\mathfrak F\vert_\mathcal O$, $\mathcal O\subset \Phi(M)$.

\section{The problem of a polynomial commutative integrability}\label{sec4}

As above, we suppose that the Euler equations
\eqref{euler-0} are completely integrable by means of a complete
commutative algebra $\mathfrak A_0$ of polynomial integrals and set $\mathcal A_0=\pr_{\g_0}^*\mathfrak A_0$.

According to Lemma~\ref{trof} and Theorem~\ref{th1}, we have

\begin{corollary}\label{cor1}
Suppose that for every $i=1,\dots,n$ there exist a commutative
subalgebra $\mathfrak B_i$ of the algebra of $\Ad_{G_{i-1}}$--invariants $\mathfrak
A_i=\R[\g_i]^{G_{i-1}}$, such that
$\mathfrak B_i+\pr_{\g_{i-1}}^*(\R[\g_{i-1}])$
is a complete algebra on $\g_i$. Then
\begin{align}\label{miki}
\mathcal B=\mathcal A_0+\mathcal B_1+\dots+\mathcal B_n, \quad \mathcal
B_i=\pr_{\g_i}^*(\mathfrak B_i), \quad i=1,\dots,n
\end{align}
is a complete commutative set on $\g$:
\begin{align*}
\ddim\mathcal B=\mathbf a(\g)=\frac12\big(\dim\g+\rank\g\big).
\end{align*}
 \end{corollary}

Therefore, the polynomial commutative integrability of the system
\eqref{euler-0},~\eqref{euler-i}, reduces to a
construction of commutative subalgebras $\mathfrak B_i$ of $\mathfrak A_i$,
 $i=1,\dots,n$.

Mikityuk proved that Bogoyavlenski's
integrals~\eqref{intB} solves the problem in the case when $(\g_i,\g_{i-1})$ are symmetric pairs~\cite{Mik} (see also examples in~\cite{Ha, PY1}).

To simplify notation, let us denote $K=G_{i-1}$, $G=G_i$, $\kk=\g_{i-1}$, $\g=\g_i$, $\mathfrak p=\mathfrak p_i$, $x=x_i$, $y_0=x_{i-1}$, $y_1=y_i$,
$\mathfrak A=\R[\g]^{K}$. The algebra of polynomials
\begin{align}\label{complete}
\mathfrak A+\pr_{\kk}^*(\R[\kk])
\end{align}
is complete  with respect to the Lie Poisson bracket on $\g$ (see \eqref{complete_i}).

Let $\mathfrak B$ be a commutative subset of the algebra of $\Ad_K$--invariants $\mathfrak A$. We always assume that $\mathfrak B$ contains
$\Ad_G$--invariant polynomials
$\R[\g]^G$ and  the lift of $\Ad_K$--invariants $\pr_{\kk}^*(\R[\kk]^K)$ that belong to the center of $\mathfrak A$.

\begin{definition}
We shall say that $\mathfrak B$ is a \emph{complete commutative set of $\Ad_K$--invariants} (or $\ad_\kk$--\emph{invariants})
if the set of polynomials
\begin{align*}
\mathfrak B+\pr_{\kk}^*(\R[\kk])
\end{align*}
is complete on $\g$ with respect to the Lie-Poisson bracket,
or, equivalently, if
$\mathfrak B$ is a complete commutative subset of $\mathfrak A$:
\begin{align*}
\ddim\mathfrak B=\dind\mathfrak B=\frac12(\ddim\mathfrak A+\dind\mathfrak A).
\end{align*}
\end{definition}

\begin{theorem}\label{th2}
 A commutative set of $\Ad_K$--invariants $\mathfrak B$ is complete if it has
\begin{align*}
\mathbf b(\g,\kk)=\frac12(\dim\mathfrak p+\rank\g-\rank\kk)
\end{align*}
polynomials independent from polynomials $\pr^*_\kk(\R[\kk])$.
In other words, $\mathfrak B$ is complete if
\begin{align*}
\dim\pr_{\mathfrak p}\Span\{\nabla p(x) \,\vert\, p\in\mathfrak B\}=\frac12(\dim\mathfrak p+\rank\g-\rank\kk),
\end{align*}
for a generic $x\in\g$.
\end{theorem}

\begin{proof}
Assume that  $\mathfrak B$ is a commutative set of $\Ad_K$--invariants and let
\begin{align*}
\kappa=\dim\pr_{\mathfrak p}\Span\{\nabla p(x) \,\vert\, p\in\mathfrak B\}.
\end{align*}

Then, due to inequality~\eqref{nejednakost}  and the completeness of \eqref{complete}, we have
\begin{align*}
(\kappa+\dim\kk)+(\kappa+\rank\kk)&=
\ddim(\mathfrak B+\pr^*_{\kk}(\R[\kk]))+\ddim(\mathfrak B+\pr^*_{\kk}(\R[\kk])) \\
&\le \ddim(\mathfrak A+\pr^*_{\kk}(\R[\kk]))+\ddim(\mathfrak A+\pr^*_{\kk}(\R[\kk])) \\
&=2\mathbf a(\g)=\dim\g+\rank\g.
\end{align*}

Therefore, $\mathfrak B+\pr^*_{\kk}(\R[\kk])$ is complete if and only if $\kappa=\mathbf b(\g,\kk)$.
\end{proof}

Note that the construction of $\mathfrak B$ is closely related to the construction of complete $G$--invariant algebras of functions on the cotangent bundle of the homogeneous space $G/K$
(see~\cite{BJ1, BJ3, MP, LP, Mik3, Jo, Jo2, DGJ, DGJ2}).

The simplest situation is the case when the algebra of $\Ad_K$--invariants $\mathfrak A$ is already commutative -- the center of $\mathfrak A$ coincides with $\mathfrak A$. Then we say that
$K\subset G$ is a \emph{multiplicity free subgroup} of $G$. For example, $SO(n-1)$ and $U(n-1)$ are multiplicity free subgroups of $SO(n)$ and $U(n)$, respectively.
This is the reason that the invariants~\eqref{thimm} are sufficient for the integrability of
the Gel'fand-Cetlin systems on $\so(n)$ and $\mathfrak{u}(n)$~\cite{GS1, GS2}.

The next natural step is to consider a subgroup $K\subset G$ when apart from $\Ad_G$--invariants, for a complete commutative set of $\Ad_K$--invariants we can take arbitrary $\Ad_K$--invariant polynomial, which is not in the center of $\mathfrak A$.
Then we say that $K$ is an \emph{almost multiplicity free subgroup} of $G$.

The classification of multiplicity free subgroups $K$ of compact Lie groups $G$ is given by Kr\"{a}mer~\cite{Kr} (see also Heckman~\cite{He}).
If $G$ is a simple group, the pair of corresponding
Lie algebras $(\g,\kk)$ is
\begin{align*}
(B_n,D_n), \quad (D_n,B_{n-1}), \quad   \text{or} \quad \text (A_n,A_{n-1}\oplus u(1)).
\end{align*}

\begin{example}\label{ex1} Multiplicity free pairs are:
\begin{align*}
&(SU(n),S(U(1)\times U(n-1))),\, (SU(n),U(n-1)), \, (SU(4),Sp(2)), \\
&(SO(n),SO(n-1)),\, (SO(4),U(2)),\, (SO(4),SU(2)),\\
& (SO(6),U(3)),\, (SO(8),Spin(7)),\,  (Spin(7),SU(4)).
\end{align*}
\end{example}

In the next statement we obtained the classifications of almost multiplicity free subgroups of compact simple Lie groups:

\begin{theorem}
The pair of Lie algebras $(\g,\kk)$ corresponding to the almost multiplicity free subgroups $K\subset G$ belongs to the following list:
\begin{align*}
&(A_n, A_{n-1}), \quad (A_3, A_1\oplus A_1\oplus\uu(1)), \quad (B_2, \uu(2)), \\
&(B_2, B_1\oplus\uu(1)),  \quad (B_3, \g_2),\quad (\g_2, A_2).
\end{align*}
\end{theorem}

The proof will be given in a separate paper.

\begin{example} \label{ex2} Almost multiplicity free pairs:
\begin{align*}
& (SU(n),SU(n-1)), \, (SU(4),S(U(2)\times U(2)), \, (SU(3),SO(3)),\\
& (SO(5),SO(3)\times SO(2)), \, (Sp(2),Sp(1)\times U(1)), \, (SO(5),U(2)), \\
& (Sp(2),U(2)),\, (SO(6),SO(4)\times SO(2)), \, (SO(6),SU(3)),\\
& (Spin(7), G_2), \, (G_2,SU(3)), \, (SO(3)\times SO(4),SO(3)).
\end{align*}
\end{example}

\section{Commutative  polynomial integrals}\label{sec6}

As in Section~\ref{sec4}, let us denote $K=G_{i-1}$, $G=G_i$, $\kk=\g_{i-1}$, $\g=\g_i$, $\mathfrak p=\mathfrak p_i$, $x=x_i$, $y_0=x_{i-1}$, $y_1=y_i$,
that is $x=y_0+y_1$.
Further, in this section by $p_1,\dots,p_{r}$ we denote the base of homogeneous invariant
polynomials on $\g$, $r=\rank\g$.

\subsection{Isotropy subgroups and the Mishcenko-Fomenko shifting of argument method}

Mishchenko and Fomenko showed that the set of polynomials induced
from the invariants by shifting the argument
\begin{align}
& {\mathfrak B}=\{p_{j,a,k}(x)\,\vert\,
k=0,\dots,\deg p_j, \, j=1,\dots,r\},\label{shift}\\
& p_{j,a,\lambda}=p_j(x+\lambda a)=\sum_{k=0}^{\deg p_j} p_{j,a,k}(x)\lambda^k
\nonumber
\end{align}
is a complete commutative  set on $\mathfrak g$, for a generic
$a\in \mathfrak g$ (see~\cite{Bo2, MF1}).

The following Bolsinov's statement will be useful to us frequently, hence we are stating it for the sake of completeness. The proof can be found in~\cite[Proposition 2.5]{Bo2}.

\begin{lemma}[\cite{Bo2}]\label{bolsinov}
{Consider a pencil of skew-symmetric linear forms
$
\Pi=\{\lambda_1\Lambda_1+\lambda_2\Lambda_2\, \vert\, \lambda_1,\lambda_2\in\R, \,\vert\lambda_1\vert+\vert\lambda_2\vert\ne 0\}
$
in $\R^n$ and set $
R_0=\max_{\Lambda\in\Pi}\rank \Lambda$.
Let
$\Lambda^1,\dots,\Lambda^q\in \Pi^\mathbb C=\{\lambda_1\Lambda_1+\lambda_2\Lambda_2\, \vert\, \lambda_1,\lambda_2\in\mathbb C, \,\vert\lambda_1\vert+\vert\lambda_2\vert\ne 0\}$ be linearly independent skew-symmetric forms in $\mathbb C^n$ with rank less then $R_0$.
Let $L\subset \mathbb C^n$ be the union of kernels of all forms in $\Pi^\mathbb C$,
$L_0\subset \mathbb C^n$ be the union of kernels of forms with the maximal rank, and let
\begin{align*}
L_0^{\Lambda}=\{\xi\in \mathbb C^n\,\vert\, \Lambda(\xi,\eta)=0, \, \eta\in L_0\}
\end{align*}
be the $\Lambda$--orthogonal space of $L_0$. Then.}

\begin{enumerate}[wide, topsep=0pt, itemsep=1pt, labelwidth=!, labelindent=0pt, label=(\roman*)]
\item $\Lambda$--orthogonal space of kernels of forms with the maximal rank $L_0^{\Lambda}$
does not depend on $\Lambda\in\Pi^\mathbb C$.
It is an isotropic subspace and contains the kernels of all forms in $\Pi^\mathbb C$:
$L\subset L_0^{\Lambda}, \quad  L\subset L_0^{\Lambda}$.

\item The equality $L_0^{\Lambda}=L$ holds if and only if
\begin{align*}
\dim_\mathbb C\ker(\Lambda\vert_{\ker \Lambda^i})=\dim\{\xi\in\ker\Lambda^i\, \vert\,\Lambda(\xi,\eta)=0, \, \eta\in\ker\Lambda^i \}=n-R_0, \quad i=1,\dots,q,
\end{align*}
where $\Lambda\in\Pi^\mathbb C$ is a skew-symmetric form of rank $R_0$ at $x$.
\end{enumerate}
\end{lemma}

\begin{theorem}\label{singMF}
Let $\kk$ be a Lie algebra equal to the isotropy
algebra of a element $a\in\g$
\begin{align*}
\kk=\g(a)=\{x\in \g\,\vert\, [x,a]=0\}
\end{align*}
and let $K$ be its corresponding Lie group.
The set~\eqref{shift} is a complete commutative set of $\Ad_K$-invariant
polynomials on $\g$.
\end{theorem}

\begin{proof} It is clear that polynomials
$\mathfrak B$ are $\Ad_K$-invariant. Based on Lemma~\ref{bolsinov}, Bolsinov proved that
for a given $x^0\in\g$ (regular or
singular), one can find a regular element $a\in\g$, such that
$\mathfrak B$ is a complete commutative algebra on the adjoint
orbit through $x^0$ (see~\cite{Br} and \cite[Theorem 5,  page 230]{TF}, the statement is firstly proved by Mishchenko and Fomenko by using a different approach~\cite{MF1, TF}).
One can reverse the roll of $x^0$ and
$a$ and slightly change the given theorem to prove the required statement.
We will present the detailed proof since some of the
arguments will be used in the proof of Theorem \ref{GCM} given below.

The set~\eqref{shift} is the union of Casimir polynomials of the Poisson brackets of maximal rank $R_0=\dim\g-\rank\g$
within the pencil $\Pi$ of compatible Poisson brackets given by Lie Poisson structure $\Lambda$ and the $a$-bracket $\Lambda_a$ given by
\begin{align}\label{aB}
\{f,g\}_a\vert_x=\Lambda_a(x)(\nabla f(x),\nabla g(x))=-\langle a,[\nabla f(x),\nabla g(x)]\rangle.
\end{align}

It is sufficient to consider the case when $a$ is a singular element of $\g$. Since $\g$ is compact, there exist $x=y_0+y_1\in\g$, such that the complex plane
\begin{align*}
\ell(x,a)=\{\lambda_0 y_{0}+\lambda_1
y_1\,\vert\,\lambda_0,\lambda_1\in\mathbb C\}
\end{align*}
intersect the set of singular points in the complexified Lie algebra $\g^\mathbb C$ only at the line $\mathbb C\cdot a$ and that $y_0$ is regular in $\kk^\mathbb C$.

Consider the complexified pencil of skew-symmetric forms
\begin{align*}
\Pi^\mathbb C_x=\{\lambda_0\Lambda(x)+\lambda_1\Lambda_a(x)\, \vert\, \lambda_0,\lambda_1\in\mathbb C, \,\vert\lambda_0\vert+\vert\lambda_1\vert\ne 0\}.
\end{align*}

The kernel of $\lambda_0\Lambda(x)+\lambda_1\Lambda_a(x)$ in $\g^\mathbb C$ is the isotropy algebra
\begin{align}\label{shift1}
\g^\mathbb C(\lambda_0x+\lambda_1 a)=\{\xi\in \g^\mathbb C\,\vert\, [\xi,\lambda_0x+\lambda_1\lambda a]=0\}.
\end{align}
Thus, all forms $\Pi^\mathbb C_x$ are regular except $\Lambda_a(x)$ with the kernel equals to $\kk^\mathbb C$.
According to Lemma~\ref{bolsinov}, we have
\begin{align}\label{shift2}
L_0^{\Lambda}=L_0+\kk^\mathbb C
\end{align}
if and only if\footnote{By $\langle\cdot,\cdot\rangle$ we also denote an invariant quadratic form on $\g^\mathbb C$, the extension of the invariant scalar product from $\g$ to $\g^\mathbb C$.}
\begin{align*}
\dim_\mathbb C\{\xi\in\kk^\mathbb C\, \vert\, \langle x,[\xi,\eta]\rangle=0, \, \eta\in\kk^\mathbb C\}=
\dim_\mathbb C\{\xi\in\kk^\mathbb C\, \vert\, [\xi,y_0]=0\}=\rank\g.
\end{align*}

Since $\rank\kk=\rank\g$ and $y_0$ is regular, the above identity is satisfied.
The real part of $L_0$ is spanned by the gradients of the polynomials in $\mathfrak B$ at $x$. Therefore,
in the real domain,~\eqref{shift2} implies that $\mathfrak B+\pr_{\kk}^*(\R[\kk])$
is a complete set of functions at $x$, and therefore on an open dense subset on $\g$.
\end{proof}

\subsection{Mikityuk's construction with symmetric pairs and its variation}

In the above notation, Bogoyavlenski's
integrals~\eqref{intB} can be defined  as coefficients in $\lambda$ in the expansion of $p_j(y_{0}+\lambda y_1)$:
\begin{align}
 \mathfrak B&=\{p_{j,k}(x)\,\vert\,
k=0,\dots,\deg p_j,\,j=1,\dots,r\}, \label{gcm}\\
p_{j,\lambda}(x)&=p_j(y_{0}+\lambda y_1)=\sum_k
\lambda^k p_{j,k}(x), \qquad j=1,\dots,r.
\nonumber
\end{align}

Mikityuk has proved the following completeness
statement for polynomials~\eqref{gcm} (see~\cite[Proposition 3,  Theorems 1 and  2]{Mik}).

\begin{theorem}[\cite{Mik}]\label{MikTh}
Assume that $(\g,\kk)$ is a symmetric pair. Then the set of polynomials~\eqref{gcm} is a complete commutative set of $\ad_\kk$--invariant polynomials on $\g$. Therefore, if all pairs $(\g_i,\g_{i-1})$, $i=1,\dots,n$
are symmetric and $\mathfrak A_0$ is a complete commutative set on $\g_0$, then the associated set~\eqref{miki} is a complete commutative set on $\g$.
\end{theorem}

There is a small variation of Mikityuk's construction that allows us to significantly extend the class of examples if not all of the pairs $(\g_i,\g_{i-1})$, $i=1,\dots,n$ are symmetric. Simply, we can extend the original filtration and use different methods at every step.
As an illustration, consider an arbitrary chain of compact symplectic groups with standard inclusions
\begin{align}\label{sp-fil}
Sp(k_0)\subset Sp(k_1)\subset \dots\subset Sp(k_n), \qquad k_0 < k_1 <\dots <k_n.
\end{align}
Then, we extend~\eqref{sp-fil} to the filtration (also using natural inclusions):
\begin{align*}
Sp(k_0)\subset Sp(k_0)\times Sp(k_1-k_0)\subset Sp(k_1)\subset  \dots\subset Sp(k_{n-1})\times Sp(k_n-k_{n-1})\subset  Sp(k_n).
\end{align*}

Now, the construction of functions in involution is clear. If in the step $(\g_i,\g_{i-1})$ we have a symmetric pair $(\spp(k_j),\spp(k_{j-1})\times \spp(k_j-k_{j-1}))$,
then $\mathcal B_i$ is given by~\eqref{gcm}. On the other hand, if $(\g_i,\g_{i-1})$ is $(\spp(k_{j-1})\times \spp(k_j-k_{j-1}),\spp(k_{j-1}))$, then
for $\mathcal B_i$ we take an arbitrary complete commutative set on $\spp(k_j-k_{j-1})$ (for example using the argument shift method~\cite{MF1}).

\begin{proposition}\label{sp-thm}
Assume that the Euler equations~\eqref{euler-0} on $\spp(k_0)$
are integrable by polynomial integrals. Then the Euler equations~\eqref{euler-0},
\eqref{euler-i} associated to the filtration~\eqref{sp-fil} are completely integrable in a commutative sense by means of polynomial integrals as well.
\end{proposition}

Of course, the above variation can be applied for other constructions of commutative polynomials.
If for a given Lie subalgebra $\kk\subset \g$ one can find a Lie subalgebra $\mathfrak h$, $\kk\subset\mathfrak h\subset \g$, having a complete commutative set of $\ad_\kk$--invariant polynomials on $\mathfrak h$ and a complete commutative set of $\ad_\mathfrak h$--invariant polynomials on $\g$, then the problem is solved for a
pair $(\g,\kk)$ as well.

\subsection{Bogoyavlenski's integrals}\label{poredjenje}

 First, we note the following lemma.
\begin{lemma}\label{GCMprop}
The polynomials~\eqref{gcm} are $\Ad_K$--invariant.
\end{lemma}
\begin{proof}  As we already mentioned,
the polynomials~\eqref{gcm} commute with the lifting of polynomials from $\R[\mathfrak k]$ to $\R[\g]$ if $(\g,\kk)$ is a symmetric pair (Theorem~\ref{MikTh}).
The proof given there can be slightly modified, and adopted to a general case.

Let $p\in \R[\g]^G$ be an invariant polynomial and let $p_\lambda(x)=p(y_0+\lambda y_1)$.
The gradient of $p_{\lambda}$ at $x$ is given by
\begin{flalign*}
&&P^\lambda&=\nabla p_{\lambda}(x)=P^\lambda_0 +\lambda P^\lambda_1,&\\
\text{where} &&
P^\lambda_0&=\pr_\kk \nabla p\vert_{y_0+\lambda y_1}, \quad P^\lambda_1=\pr_\mathfrak p \nabla p\vert_{y_0+\lambda y_1}.&
\end{flalign*}
Since $p$ is an invariant polynomial,  we have
\begin{align}\label{InvPi}
[P^{\lambda}_0+P^\lambda_1, y_{0}+\lambda y_1]=0.
\end{align}

Let $\xi$ be an arbitrary element in $\kk$. Then
\begin{align}
\langle \nabla p_{\lambda}(x),[\xi,x]\rangle = \langle P_0^\lambda +\lambda P_1^\lambda,[\xi,y_0+y_1]\rangle
= \langle P_0^\lambda,[\xi,y_0]\rangle+ \lambda \langle  P_1^\lambda,[\xi,y_1]\rangle. \label{InvPi2}
\end{align}
On the other hand, from
\eqref{InvPi} we get
\begin{align}\label{InvPi3}
0=\langle [P_0^\lambda +P_1^\lambda,y_0+\lambda y_1],\xi\rangle=\langle [P_0^\lambda,y_0],\xi\rangle+\langle [P_1^\lambda,\lambda y_1],\xi\rangle.
\end{align}
Therefore, according to~\eqref{InvPi2} and~\eqref{InvPi3}, $p_{\lambda}$ is an $\Ad_K$--invariant polynomial.

We can see this also directly by using the identities: $p(\Ad_g(x))=p(x)$, $\Ad_g(x)=\Ad_g(y_0)+\Ad_g(y_1)$,  $g\in G$,  and $\Ad_g(y_0)\in\kk$, $\Ad_g(y_1)\in\mathfrak p$, $g\in K$.
\end{proof}

Now, we would like to use a relationship between the argument shift method and translations along the subalgebras presented above to estimate the number of independent Bogoyavlenski's integrals.

\begin{theorem}\label{GCM}
Let $Sing(\g^\mathbb C)$ be the set of singular point in $\g^\mathbb C$ and let
\begin{align*}
\ell(x)=\{\lambda_0 y_{0}+\lambda_1
y_1\,\vert\,\lambda_0,\lambda_1\in\mathbb C\}.
\end{align*}
Assume that there exist $x\in\g$ such that
\begin{align*}
  \ell(x)\cap Sing(\g^\mathbb C)=\{0\},\qquad \text{or,}\qquad
 \ell(x)\cap Sing(\g^\mathbb C)=\mathbb C\cdot y_1.
\end{align*}
Then, for a given $x$, for polynomials~\eqref{gcm} we have
\begin{align*}
\dim B(x)\ge \mathbf b(\g,\kk)=\frac12(\dim\mathfrak p+\rank\g-\rank\kk),
\end{align*}
where
$B(x)=\pr_{\mathfrak p}\Span\{\nabla p_{j,\lambda}(x)\,\vert\,j=1,\dots,r,\,\lambda\in\R\}.$
\end{theorem}
\begin{proof} \emph{Step 1.}
  We can obtain polynomials~\eqref{gcm} by the translation of invariants in the direction of $\kk$ instead of $\mathfrak p$: $p_\lambda(x)=p(\lambda y_0+y_1)$, $p\in\R[\g]^G$.
Since
\begin{align*}
\nabla p_{\lambda}(x)=\lambda \pr_{\mathfrak k}\nabla
p\vert_{\lambda y_{0}+ y_1}+\pr_{\mathfrak
p}\nabla p\vert_{\lambda y_{0}+y_1}=\lambda \pr_{\mathfrak k}\nabla
p\vert_{\lambda y_{0}+ y_1}+\pr_{\mathfrak
p}\nabla p\vert_{{x+\mu y_{0}}},
\end{align*}
$\lambda=\mu+1$,
we need to estimate the dimension of the linear space
\begin{align*}
B(x)=\Span\{\pr_{\mathfrak p}\nabla p_j\vert_{x+\mu
y_{0}}\,\vert\, j=1,\dots,r,\,\mu\in\R\}.
\end{align*}
The space $B(x)$ is equal to the projection to $\mathfrak p$
of the space $C(x)$ spanned by gradients of the polynomials
$p_{a,\mu}(x)=p(x+\mu a)$
obtained by shifting of argument in the direction
$a=y_{0}=\pr_{\kk}(x)$:
\begin{align*}
C(x)=\Span\{\nabla p_{a,\mu}(x)\,\vert\,p\in\R[\g]^G\}
\end{align*}
Note that the sets of polynomials~\eqref{shift} and~\eqref{gcm} are different, but the projections to $\mathfrak p$
of their gradients at the given point $x$ are the same.

\

\emph{Step 2.}
Let us assume that the complex plane $\ell(x)$ intersects the
set of singular points in $\g^\mathbb C$ only in $0$ and consider the pencil of compatible Poisson structures
spanned by Lie-Poisson bracket~\eqref{LP} and the $a$-bracket~\eqref{aB}, where $a=y_0$.

The kernel of the skew-symmetric form $\lambda_0\Lambda(x)+\lambda_1\Lambda_a(x)$ in $\g^\mathbb C$ is the isotropy algebra~\eqref{shift1}.
Thus all forms in $\Pi^\mathbb C_x$ have the maximal rank, and, according to Lemma~\ref{bolsinov},
{$L_0^\Lambda=L_0$ where $L_0\subset\g^\mathbb C$ is the union of all kernels in in $\Pi^\mathbb C_x$. Since $C(x)=L_0\cap \g$,
the set $\{p_{a,\mu}\,\vert \,\mu\in\R\}$} is complete at $x$:
\begin{align*}
\dim C(x)=\mathbf a(\g)=\frac12(\dim\g+\rank\g).
\end{align*}

Therefore
\begin{align*}
\dim B(x)=\dim\pr_{\mathfrak p} C(x)=
\frac12(\dim\g+\rank\g)-\dim(C(x)\cap \kk).
\end{align*}

It remains to note that
\begin{align*}
\dim(C(x)\cap \kk)\le \frac12(\dim\kk+\rank\kk),
\end{align*}
implying that $\dim B(x)\ge \mathbf b(\g,\kk)$.
Indeed, we have
\begin{align*}
\{p_{a,\mu_1}, f_{a,\mu_2}\}(x)=
\langle x,[\nabla p\vert_{x+\mu_1 a},\nabla
f\vert_{x+\mu_2 a}]\rangle=0, \qquad p,f\in\R[\g]^G,
\end{align*}
and, if $\nabla p\vert_{x+\mu_1 a},\nabla
f\vert_{x+\mu_2 a} \in\kk$, then also
\begin{align*}
\langle y_{0},[\nabla p\vert_{x+\mu_1 a},
\nabla f\vert_{x+\mu_2 a}]\rangle=0.
\end{align*}

Thus, $C(x)\cap \kk$ is an isotropic subspace of $\kk$ at
$y_{0}$ with respect to the Lie-Poisson bracket on $\kk$. On the other
hand, the maximal isotropic subspace at $y_{0}$ (it is regular
in $\kk$) has the dimension
$\frac12(\dim\kk+\rank\kk)$.

\

\emph{Step 3.} Now assume that the complex plane $\ell(x)$ intersects the
set of singular points in $\mathbb C\cdot y_1$ and in addition that
$\pr_{\g^\mathbb C(y_1)} x$
is regular in $\mathfrak g^\mathbb C(y_1)$,
where $\mathfrak g^\mathbb C(y_1)$ is the isotropy Lie algebra
of $y_1$ within $\mathfrak g^\mathbb C$.
Then all skew-symmetric forms in $\Pi^\mathbb C_x$, except $\Lambda^1=\Lambda(x)-\Lambda_a(x)$,
 have the maximal rank $(a=y_0)$.
Since $\rank{\g^\mathbb C(y_1)} =\rank\g$, we get
\begin{align*}
\dim_\mathbb C\ker(\Lambda\vert_{\ker \Lambda^1}) =&
\dim_\mathbb C\{\xi\in\g^\mathbb C(y_1)\, \vert\, \langle x,[\xi,\eta]\rangle=0, \, \eta\in\g^\mathbb C(y_1)\}\\
=&\dim_\mathbb C\{\xi\in\g^\mathbb C(y_1)\, \vert\, [\xi,\pr_{\g^\mathbb C(y_1)} x]=0\}=\rank\g.
\end{align*}

Thus, from Lemma~\ref{bolsinov}, we have
$L_0^{\Lambda}=L_0+\g^\mathbb C(y_1)$,
that in the real domain implies
\begin{align}\label{shift4}
C(x)^\Lambda=C(x)+\g(y_1).
\end{align}

For $y_1$ that belong an open dense set of $\mathfrak p$ with minimal dimensions of the isotropy algebras
$\g(y_1)$ and $\kk(y_1)$, we have that the semisimple part of $\g(y_1)$ belongs to $\kk(y_1)$ (e.g., see~\cite{Mik3}):
\begin{align*}
[\g(y_1),\g(y_1)]=[\kk(y_1),\kk(y_1)].
\end{align*}
Then
\begin{align}\label{ann}
\g(y_1)=\kk(y_1)+\mathfrak z(\g(y_1)),
\end{align}
where $\mathfrak z(\g(y_1))$ is the center of $\g(y_1)$. Further, we have
(e.g., see ~\cite[Lemma 4]{Baz})
\begin{align}\label{ann*}
\mathfrak z(\g(y_1))=\Span\{\nabla p_j(y_1)\, \vert\, j=1,\dots,r\} \subset C(x).
\end{align}

By combining~\eqref{shift4},~\eqref{ann}, and~\eqref{ann*} we obtain
\begin{align*}
C(x)^\Lambda=C(x)+\kk(y_1).
\end{align*}
Thus, there exist a subspace $D\subset \kk(y_1)$ such that $C_1(x)=C(x)+D$ is a maximal isotropic subspace in $\g$ with respect to $\Lambda(x)$ and
\begin{align*}
\dim C_1(x)=\frac12(\dim\g+\rank\g).
\end{align*}

Using the identity $B(x)=\pr_\mathfrak p C(x)=\pr_\mathfrak p C_1(x)$,
the rest of the proof is the same as in Step 2 with $C(x)$ replaced by $C_1(x)$.
\end{proof}

{Thus, if $\mathfrak B$ is commutative and the conditions of Theorem~\ref{GCM} are satisfied, according to Theorem~\ref{th2} and Lemma~\ref{GCMprop},  \eqref{gcm}~is a complete commutative set of $\ad_\kk$--invariant polynomials on $\g$.

Let $f,p$ be invariant polynomials, $f_\lambda(x)=f(y_0+\lambda y_1)$, and $p_\mu(x)=p(y_0+\mu y_1)$, $\mu\ne\lambda$, $\lambda^2+\mu^2\ne 0$. With the above notation we have
\begin{align}\label{kom1}
[F^{\lambda}_0+F^\lambda_1, y_{0}+\lambda y_1]=0, \qquad [P^{\mu}_0+P^\mu_1, y_{0}+\mu y_1]=0,
\end{align}
which implies (see~\eqref{InvPi3}):
\begin{align*}
\langle y_0, [F_0^\lambda,\xi]\rangle= \lambda \langle y_1, [\xi, F_1^\lambda]\rangle, \quad \langle y_0, [P_0^\mu,\xi]\rangle= \mu \langle y_1, [\xi, P_1^\mu]\rangle, \quad \xi\in\kk.
\end{align*}
In particular,
\begin{align}\label{kom2}
\langle y_0, [F_0^\lambda,P_0^\mu]\rangle= \lambda \langle y_1, [P^\mu_0, F_1^\lambda]\rangle, \quad \langle y_0, [P_0^\mu,F^\lambda_0]\rangle= \mu \langle y_1, [F^\lambda_0, P_1^\mu]\rangle.
\end{align}
Since $x=y_0+y_1$, $y_0$, $y_1$ can be expressed as linear combinations of $y_0+\lambda y_1$ and $y_0+\mu y_1$, from~\eqref{kom1} and~\eqref{kom2} we get
\begin{align*}
0=\langle y_0, [P_0^\mu+P^\mu_1,F^\lambda_0+F^\lambda_1]\rangle=&\langle y_0, [P_0^\mu,F^\lambda_0]\rangle+\langle y_0, [P^\mu_1,F^\lambda_1]\rangle,\\
0=\langle y_1, [P_0^\mu+P^\mu_1,F^\lambda_0+F^\lambda_1]\rangle=&\langle y_1, [P_0^\mu,F^\lambda_1]\rangle+\langle y_1, [P^\mu_1,F^\lambda_0]\rangle+
              \langle y_1, [P^\mu_1,F^\lambda_1]\rangle\\
          =&\big(\frac1\lambda+\frac1\mu\big)\langle y_0, [F_0^\mu,P^\lambda_0]\rangle+\langle y_1, [P^\mu_1,F^\lambda_1]\rangle.
\end{align*}
Now, by the use of the above identities, the Poisson bracket $\{f_\lambda,p_\mu\}$ reads
\begin{align*}
\{f_\lambda,p_\mu\}\vert_x=&\langle y_0+y_1, [P_0^\mu+\mu P^\mu_1,F^\lambda_0+\lambda F^\lambda_1]\rangle\\
=&\langle y_0, [P_0^\mu,F^\lambda_0]\rangle+\lambda\mu\langle y_0, [P^\mu_1,F^\lambda_1]\rangle\\
&+\lambda\langle y_1, [P_0^\mu,F^\lambda_1]\rangle+\mu\langle y_1, [P^\mu_1,F^\lambda_0]\rangle+
\lambda\mu\langle y_1, [P^\mu_1,F^\lambda_1]\rangle\\
=&\langle y_0, [P_0^\mu,F^\lambda_0]\rangle+\lambda\mu\langle y_0, [F_0^\mu,P^\lambda_0]\rangle\\
&+\langle y_0, [F_0^\mu,P^\lambda_0]\rangle+\langle y_0, [F_0^\mu,F^\lambda_P]\rangle-
(\lambda+\mu)\langle y_0, [F_0^\mu,P^\lambda_0]\rangle\\
=&(1-\lambda)(1-\mu)\langle y_0, [F_0^\mu,P^\lambda_0]\rangle.
\end{align*}
}

\begin{example}
If $\kk$ is a Cartan subalgebra, then $\{f_\lambda,p_\mu\}\vert_x=(1-\lambda)(1-\mu)\langle y_0, [F_0^\mu,P^\lambda_0]\rangle=0$ and
integrals~\eqref{gcm} provides a complete commutative set on $\g$. The pair $(\g,\kk)$ is already described in Theorem \ref{singMF}: $\kk=\g(a)$, where $a\in\kk$
is regular.
\end{example}

It is also obvious that if $(\mathfrak g,\kk)$ is a symmetric pair that $\mathfrak B$ is commutative.
 In Bogoyavlenski~\cite[Theorem 1]{B1} it is claimed that the set $\mathfrak B$ is always commutative,
 however the presented proof contains a small gap. Recently, Panyushev and Yakimova gave an example of the case where $\mathfrak B$ is not commutative~\cite[Example 2.3]{PY3}.

\begin{example}
For symmetric pairs $(\so(n),\so (p)\times \so (n-p))$, Theorem~\ref{GCM} provides another proof of Theorem~\ref{MikTh}.
\end{example}

\begin{example}\label{exCom}
In the following example we verified the commutativity of $\mathfrak B$
by direct computations.
The examples are also convenient in the discussion of the conditions in Theorem~\ref{GCM}.
Consider the case $\g=\so(5)$. The Lie subalgebras
\begin{align*}
\so(2)\oplus\so(2), \quad \so(2)\oplus\so(3), \quad  \so(4)
\end{align*}
satisfy conditions of Theorem~\ref{GCM}. Moreover, $SO(4)$ and $SO(2)\times SO(3)$ are multiplicity free and almost multiplicity free subgroups of $SO(5)$.
On the other side, if $\kk=\so(3)$ or $\kk=\so(2)$, a generic $y_0\in\kk$ is not regular in $\so(5)$ and we can not apply Theorem~\ref{GCM}.

We have
\begin{align}\label{USLOV}
\mathbf{b}(\so(5),\so(3))=4, \qquad \mathbf{b}(\so(5),\so(2))=5.
\end{align}

 Let $x=y_0+ y_1\in\so(5)$, with
\begin{align*}
  y_0 &= \begin{pmatrix} 0 & 0\\ 0 & Q  \end{pmatrix},\quad
  y_1 = \begin{pmatrix} P_2 & P_1\\ -P_1^T & 0 \end{pmatrix},
\end{align*}
where $P_1\in\R^{3\times 2}$, $P_2\in\so(2)$, $Q\in \so(3)$ in the case $\kk=\so(3)$ and $P_1\in\R^{2\times 3}$, $P_2\in\so(3)$, $Q\in\so(2)$ in the case $\kk=\so(2)$ .
Bogoyavlenski's integrals are given by:
\begin{align*}
p_{1,\lambda}(x)&=p_{1,\lambda}(y_0+\lambda y_1)=\tr (y_0+\lambda y_1)^2=\sum_{k=0}^2 \lambda^k p_{j,k}(x),
\\[-10pt]
p_{1,0}(x)&= \tr(y_0^2),\\
p_{1,1}(x)&= \tr(y_0y_1 + y_1 y_0) \equiv 0,\\
p_{1,2}(x)&= \tr(y_1^2),\\
p_{2,\lambda}(x)&=p_{2,\lambda}(y_0+\lambda y_1)=\tr (y_0+\lambda y_1)^4=\smash{\sum_{k=0}^4 \lambda^k p_{j,k}(x)},
\\
p_{2,0}(x)&= \tr(y_0^4),\\
p_{2,1}(x)&= \tr(y_0y_1y_0^2 + y_1y_0^3 + y_0^2 y_1 y_0 + y_0^3 y_1)\equiv 0,\\
p_{2,2}(x)&= \tr (y_0^2 y_1^2+y_0 y_1 y_0 y_1 + y_0 y_1^2 y_0 + y_1 y_0^2 y_1 + y_1 y_0 y_1 y_0 + y_1^2 y_0^2 ),\\
p_{2,3}(x)&= \tr (y_0 y_1^3 + y_1 y_0 y_1^2 + y_1^2 y_0 y_1 + y_1^3 y_0),\\
p_{2,4}(x)&= \tr (y_1^4).
\end{align*}

The polynomials $p_{j,k}$ commute and we need to estimate the number of independent gradients after projections onto $\mathfrak p$, at a general point $x\in\so(5)$.
For this reason, we consider the gradients:
\begin{align*}
  \nabla p_{1,2}(x) &=2 y_1,\\
  \nabla p_{2,2}(x) & = 4 (y_0^2 y_1 + y_0 y_1 y_0 + y_1 y_0^2 + y_1^2 y_0 + y_1 y_0 y_1 + y_0 y_1^2),\\
  \nabla p_{2,3}(x) & = 4 (y_1^3 + y_0 y_1^2 + y_1 y_0 y_1 + y_1^2 y_0),\\
  \nabla p_{2,4}(x) & = 4 y_1^3.
\end{align*}

Let $\kk=\so(3)$ and let $e_{ij}$ be the standard basis of $\so(5)$.
 The equation
\begin{align*}
  \mu_1 \pr_{\mathfrak p}\nabla p_{1,2}(x)+\mu_2 \pr_{\mathfrak p}\nabla p_{2,2}(x)+\mu_3 \pr_{\mathfrak p}\nabla p_{2,3}(x) + \mu_4 \pr_{\mathfrak p}\nabla p_{2,4}(x)= 0
\end{align*}
at the point $x=e_{45}+e_{13}+e_{23}+e_{24}+e_{15}+e_{25}$ reduces to the system of equations
\begin{align*}
  \mu_2+\mu_3  =& 0, &
  \mu_1-8 \mu_3-8 \mu_4 =& 0, & \mu_2 = &0,\\
  \mu_3 - \mu_4 =& 0, &
  \mu_1-10 \mu_3-10 \mu_4 = &0,
\end{align*}
which has only the trivial solution. Therefore, $\mathfrak B$ is a complete set of $\Ad_{SO(3)}$--invariant polynomials.

Obviously, according to~\eqref{USLOV}, Bogoyavlenski's integrals are not sufficient for $\kk=\so(2)$.
However, we can consider the variation of the method,  by taking the chain of subalgebras
\begin{align*}
\so(2)\subset \so(2)\oplus\so(2)\subset \so(5) \quad \text{or}\quad \so(2)\subset \so(2)\oplus\so(3)\subset \so(5).
\end{align*}
\end{example}

\subsection{Diagonal subgroups}

Let $G_0$ be a compact connected Lie group and $\g_0=Lie(G_0)$ its Lie algebra.
Consider the case when the group $G$ is the product
$
G=G_0^{m}
$
and the subgroup $K\subset G$ is $G_0$ diagonally embedded into the product:
\begin{align*}
K=\diag(G_0)=\{(g,\dots,g)\,\vert\, g\in G_0\}\subset G.
\end{align*}

For a purpose of a construction of
integrable geodesic flows on $m$--symmetric spaces $Q=G/K$, the following construction related to the filtration
\begin{align*}
\g_0\subset \g_1=\g_0\oplus\g_0\subset\dots\subset
\g_{m-1}=(\g_0)^{m}=\g
\end{align*}
is given in~\cite{Jo}.
Let $f_1,\dots,f_{r_0}$ be the base of homogeneous invariant
polynomials on $\g_0$, $r_0=\rank\g_0$, and let
\begin{align*}
&\mathfrak B=\mathfrak B_1+\mathfrak B_2+\dots+\mathfrak
B_{m-1}, \\
&\mathfrak B_i=\{f^i_{\alpha,k}(x)\,\vert\,k=0,\dots,\deg
f_j, \,j=1,\dots,r_0\},
\end{align*}
where the polynomials $f^i_{j,k}(x)$ are defined by:
\begin{align*}
f_j(y_0+y_1+\dots+y_{i-1}+\lambda y_{i})=\sum_{k=0}^{\deg
f_j} f^i_{j,k}(y_0,y_1,\dots,y_{m-1})\lambda^k.
\end{align*}

\begin{theorem}[\cite{Jo}]\label{pod1}
\begin{enumerate}[wide, topsep=0pt, itemsep=1pt, labelwidth=!, labelindent=0pt, label=(\roman*)]
\item The set $\mathfrak B$ is a commutative set of
$\Ad_K$-invariant polynomials on $\g$.

\item The set $\mathcal B+\mu^*(\R[\mathfrak g_0])$ is a
complete set of polynomials on $\g$, where
\begin{align*}
\mu(y_0,y_1,\dots,y_{m-1})=y_0+y_1+\dots+y_{m-1}.
\end{align*}
\end{enumerate}
\end{theorem}

Therefore, the set $\mathfrak B$ solves our problem for the pair $(G,K)=(G_0^m,\diag(G_0))$.

\subsection{Reyman's construction: the shifting of argument and symmetric pairs}

Here we present Reyman's construction of commutative polynomials related to symmetric pairs~\cite{Re}.
Suppose that $\g$ is a Lie subalgebra of a
semisimple Lie algebra $\mathfrak h$, such that $(\mathfrak h,\g)$ is a
symmetric pair:
\begin{align*}
[\g,{\mathfrak m}]\subset {\mathfrak m}, \qquad [{\mathfrak m},
{\mathfrak m}]\subset \mathfrak g,
\end{align*}
where ${\mathfrak m}$ is the orthogonal complement of $\g$ with
respect to an invariant scalar product $\langle\cdot,\cdot\rangle$.\footnote{Again, we use the same symbol for different objects.
The restriction of $\langle\cdot,\cdot\rangle$ to
$\g$ is a positive definite invariant scalar product we are dealing with. Also, as above, we identify $\mathfrak h$ and $\mathfrak h^*$ by $\langle\cdot,\cdot\rangle$.}

Further, suppose there exist $a\in\mathfrak m$, such that
$\kk$ equals to the isotropy algebra of $a$ within~$\g$,
\begin{align*}
\kk=\g(a)=\{\, \xi\in \g \, \mid \, [\xi,a]=0\, \}.
\end{align*}

Let $h_1,\dots,h_{s}$ be the basic homogeneous
invariant polynomials on $\mathfrak h$, $s=\rank\mathfrak h$ and let $\mathcal K$ be the set of linear functions on $\kk$, considered as linear functions on $\mathfrak h$.

On $\mathfrak h$ we have a pencil $\Pi$ of
compatible Poisson bivectors spanned by
\begin{align*}
&\Lambda_1(\xi_1+\eta_1,\xi_2+\eta_2)\vert_{z}=
-\langle z, [\xi_1,\xi_2]+[\xi_1,\eta_2]+[\eta_1,\xi_2]\rangle,\\
&\Lambda_2(\xi_1+\eta_1,\xi_2+\eta_2)\vert_z=-\langle
z+a,[\xi_1+\eta_1,\xi_2+\eta_2]\rangle,
\end{align*}
where $z\in \mathfrak h$, $\xi_1,\xi_2\in \g$, $\eta_1,\eta_2\in
\mathfrak m$ (see Reyman~\cite{Re}). The Poisson bivectors $\Lambda_{\lambda_1,\lambda_2}$, for
$\lambda_1+\lambda_2\ne 0$ and  $\lambda_2 \ne 0$, are isomorphic
to the canonical Lie-Poisson bivector on $\mathfrak h$. Thus, the union of their Casimir functions
\begin{align}\label{central}
\mathcal B=\{h_{\lambda,k}(z)=h_k(\lambda x+ t+\lambda^2 a)\,
\vert\, k=1,2,\dots,s,\, \lambda\in \mathbb{R}\},
\end{align}
where $z=x+t$, $x\in \g$, $t\in \mathfrak m$, is a commutative
set with respect to the all brackets from the pencil $\Pi$
(see~\cite{Re, Bo}). Moreover, for a generic $a\in\mathfrak m$,
the set of functions $\mathcal B+\mathcal K$ is a complete
non-commutative set on $\mathfrak h$ with respect to $\Lambda_1$
(see~\cite[Theorem 1.5]{Bo}, for the detail proofs of the above
statements, given for an arbitrary semi-simple symmetric pair, see~\cite[pages 234-237]{TF}).

The symplectic leaf within $\mathfrak h$ (and the corresponding symplectic structure) of the bracket $\Lambda_1$ at a point $x\in\g$ coincide with the symplectic leaf
through $x$ of the Lie-Posson bracket (and the corresponding symplectic structure) on $\g$. Therefore, the following statement holds.
\begin{proposition}[\cite{Re, Bo2, MF1}]
The restrictions of the polynomial~\eqref{central}
\begin{align*}
& {\mathfrak B}=\{h_{j,a,k}(x)\,\vert\,
k=1,\dots,\deg f_j, \, j=1,\dots,s\},\\
& h_j(x+\lambda a)=\sum_{k=0}^{\deg h_j} h_{j,a,k}(x)\lambda^k
\end{align*}
is a commutative set of
$K$-invariant polynomials on $\g$.
\end{proposition}

Moreover, the following is true (see the discussion in~\cite[before Theorem 1.6]{Bo2}).
\begin{theorem}[\cite{Bo2}]\label{pod2}
For a generic $a\in \mathfrak m$, the
the set $\mathfrak B+\pr_{\kk}^*(\R[\kk])$ is complete on
$\g$.
\end{theorem}

Here, $a\in\mathfrak m$ is \emph{generic} if the dimension of the isotropy algebras $\g(a)$ and $\mathfrak h(a)$ are minimal.
It would be interesting to prove the above statement in the singular case
as well.
In particular, we have the following statement  (see~\cite[pages 241--244]{TF} and~\cite[Theorem 1]{DGJ}).

\begin{theorem}[\cite{TF,DGJ}]\label{pod3}
Consider the symmetric pair $(\mathfrak h,\mathfrak g)=(\mathfrak{sl}(n),\so(n))$ and let
\begin{align*}
a=\diag(\overset{n_1}{\overbrace{a_1,\dots,a_1}},\dots,\overset{n_k}{\overbrace{a_{k},\dots,a_{k}}}), \qquad \kk=\so(n)(a)=\so(n_1)\oplus\dots\oplus\so(n_k).
\end{align*}
The set $\mathfrak B$ is a complete commutative set of $\ad_\kk$--invariant polynomials on
$\so(n)$
\end{theorem}

Integrals $\mathfrak B$ are referred as Manakov integrals \cite{Ma}.  Theorem \ref{pod3} implies  complete integrability of a motion of a symmetric rigid body about a fixed point in $\R^n$ (see~\cite{Ma, FK}), having the operator of inertia $I=A^{-1}$ of the form
\[
x=I(\omega)=J\omega+\omega J,\qquad \omega\in \so(n),
\]
where a mass tensor is  $J=\diag(b_1,\dots,b_n)$,  $a_i=b_i^2$  (see~\cite[Subsection 1.6]{DGJ}).

\subsection*{Acknowledgments}
 We are grateful to the referees for thorough review and constructive comments that greatly improved quality of the paper.
The research is supported by the Project  7744592 MEGIC ”Integrability
and Extremal Problems in Mechanics, Geometry and Combinatorics” of the Science Fund
of Serbia.

\end{document}